\documentclass[showpacs,twocolumn,prb]{revtex4}%
\usepackage{amsfonts}
\usepackage{amsmath}
\usepackage{amssymb}
\usepackage{graphicx}%
\begin{document}
\title{Quantum tunneling of two coupled single-molecular magnets}
\author{Jian-Ming Hu$^1$, Zhi-De Chen$^{1,2}$ and Shun-Qing Shen$^1$}
\affiliation{$^1$Department of Physics, The University of Hong Kong, Pokfulam Road, Hong Kong, China\\
$^2$Department of Physics and Institute of Modern Condensed Matter Physics, Guangzhou University, Guangzhou 510405, China}
\begin{abstract}
Two single-molecule magnets are coupled antiferromagnetically to form a
supramolecule dimer. We study the coupling effect and tunneling process by
means of the numerical exact diagonalization method, and apply them to the
recently synthesized supramoleculer dimer [Mn$_{4}$]$_{2}$. The model
parameters are calculated for the dimer based on the tunneling process. The
absence of tunneling at zero field and sweeping rate effect on the step height
in the hysterisis loops are understood very well in this theory.

\end{abstract}
\date[October 2002]{}
\maketitle

Nanometer-sized magnetic particles and clusters have generated continuous
interests as study of their properties has proved to be scientifically and
technologically very
challenging.\cite{Awschalom95,Stamp92,Chun,Sessoli93,w1,book} Up to now
magnetic molecular clusters have been the most promising candidates for
observing quantum phenomena since they have a well defined structure with well
characterized spin ground state and anisotropy.\cite{Wernsdorfer01} One of the
well studied systems is single-molecule magnet (SMM) [Mn$_{4}$O$_{3}$%
Cl(O$_{2}$CCH$_{3}$)$_{3}$(dbm)$_{3}$] (short for Mn$_{4}$%
).\cite{Wang96,Aubin98,Andres00,Wernsdorfer02b} The molecule has well isolated
ground state with a half integer spin $S=9/2$, and magnetization tunneling is
observed at zero magnetic field. Very recently a supramolecular dimer of two
SMMs [Mn$_{4}$]$_{2}$ was reported to be synthesized
successfully.\cite{Wernsdorfer02} The antiferromagnetic coupling between two
SMMs leads to this dimer with a spin singlet ground state and makes the
quantum tunneling quite different from SMMs Mn$_{4}$. The coupling also makes
this dimer an excellent candidate for studying quantum tunneling in a system
of two truly and coupled identical particles. Quantum tunneling of
magnetization can be advantage for some applications of SMMs in providing
quantum superpositions of states required for quantum
computing.\cite{Leuenberger01, Zhou02} So the coupling effect in quantum
tunneling between SMMs is a very important issue in application of integrated
molecular magnets. In this paper we first study the tunneling process in one
SMM Mn$_{4}$ with spin $S=9/2$. A local stray field has to be introduced to
explain the tunneling of the ground state at zero field.\cite{Wernsdorfer02b}
Then we study the coupling effect of two SMMs and observe a novel triangle
tunneling process. We apply our observation to the newly synthesized
supramolecular dimer of two SMM [Mn$_{4}$]$_{2}$ and make use of the two
triangle tunneling processes to deduce the the model parameters from the
experimental data, and, furthermore, explain the sweeping rate effect in the
derivatives of hysteresis loops.

We first start with a biaxial model for a SMM Mn$_{4}$ with spin $S=9/2$,%
\begin{equation}
H_{i}=-D\mathbf{S}_{zi}^{2}+E(\mathbf{S}_{xi}^{2}-\mathbf{S}_{yi}^{2}%
)+g\mu_{B}\mu_{0}\mathbf{S}_{i}\cdot\left(  \mathbf{B}+\mathbf{h}\right)
\label{single}%
\end{equation}
where $i=1$ or $2$ referring to the two SMMs in the dimer, D and E are the
axial anisotropic constants, $\mathbf{B}=Be_{z}$ is the external magnetic
field along the z axis. The term $g\mu_{B}\mu_{0}\mathbf{S}_{i}\cdot
\mathbf{h}$ is the local stray field interaction between the SMMs and the
environment. For simplicity we denote the energy eigenstate of the biaxial
model $\left\vert m\right\rangle $ by its dominant $S_{z}=m$ component, and
$m=-S,-S+1,\cdots,S.$ The $E$ term and the stray field may lead to some minor
correction to these states. If the stray field is not included, it is well
known that for a half integer spin the tunneling between the state $\left\vert
-S\right\rangle $ and $\left\vert S\right\rangle $ is quenched due to the
parity symmetry.\cite{Hemmen86,Loss93,DiVincenzo94} It can be proved simply
that, for any integer n, we always have $\left\langle -S\right\vert \left(
H_{i}\right)  ^{n}\left\vert M\right\rangle =0$ with $M=-S+1,-S+3,\cdots,S.$
The equality indicates that there is no connection or no tunneling occurs
between these states $\left\vert -S\right\rangle $ and $\left\vert
M\right\rangle .$ Experimentally quantum tunneling was observed in a SMM
Mn$_{4}$ between the states $\left\vert -S\right\rangle $ and $\left\vert
S\right\rangle $.\cite{Aubin98} So there must be a small internal magnetic
field by the nuclear spins of the Mn ions and/or the dipole-dipole interaction
between different molecules. We model the interaction as a local stray field
$\mathbf{h}$ with a random Gaussian distribution with the equal width $\sigma$
in three directions as we did for the molecular magnets Fe$_{8},$%
\cite{Chen02}
\begin{equation}
P(\mathbf{h})=\frac{1}{(2\pi\sigma^{2})^{3/2}}\exp[-\mathbf{h}^{2}/2\sigma
^{2}].
\end{equation}
A transverse component of such a field may lead to a tunneling splitting at
zero field as observed in Ref.\cite{Aubin98}. In this paper we take the
parameters for a SMM Mn$_{4}$ $D=0.762$K, $E=0.0317$K,\cite{Aubin98,Andres00}
and $\sigma=0.035$T.\cite{Wernsdorfer02b} The resulting tunneling splitting
for the ground states $\left\vert -9/2\right\rangle $ and $\left\vert
9/2\right\rangle $ at zero field is $\sqrt{\left\langle \Delta_{0}%
^{2}\right\rangle }=3.280\times10^{-7}$K, and that for the states $\left\vert
-9/2\right\rangle $ and $\left\vert 7/2\right\rangle $ is $\sqrt{\left\langle
\Delta_{1}^{2}\right\rangle }=1.52627\times10^{-5}$K by using the exact
diagonalization method where $\left\langle \cdots\right\rangle $ stands for
the averaging over the stray field.\cite{high} Thus the local stray field may
cause a tunneling splitting between the ground states.%
\begin{figure}
[ptb]
\begin{center}
\includegraphics[
natheight=3.282800in,
natwidth=3.966900in,
height=2.0903in,
width=2.5218in
]%
{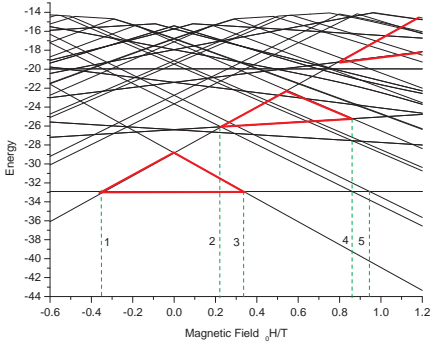}%
\caption{Spin state energy of the model Hamiltonian (Eq. (\ref{dimer})) for
[Mn$_{4}$]$_{2}$ without local stray fields as a function of applied magnetic
field by taking $D=0.762$K, $E=0.0317$K, and $J=0.1$K. The two triangles are
related to the tunneling (1) from $\left\vert -9/2,-9/2\right\rangle _{+}$ to
$\left\vert -9/2,9/2\right\rangle _{+}$ at point 1, then to $\left\vert
9/2,9/2\right\rangle _{+}$ at point 3, and (2) from $\left\vert
-9/2,-9/2\right\rangle _{+}$ to $\left\vert -9/2,7/2\right\rangle _{+}$ at
point 2, then to $\left\vert 7/2,7/2\right\rangle _{+}$ at point 4. The Point
5 is from $\left\vert -9/2,9/2\right\rangle _{+}$ to $\left\vert
9/2,7/2\right\rangle _{+}.$ The resonance fields for the five points are
$-0.335$T, $0.233$T, $0.335$T, $0.861$T, and $0.943$T, respectively. }%
\end{center}
\end{figure}

Following Wernsdorfer et al., the two SMM Mn$_{4}$s in the dimer [Mn$_{4}%
$]$_{2}$ are coupled via a weak antiferromagnetic superexchange coupling $J$.
Thus the model Hamiltonian for the dimer is
\begin{equation}
H=H_{1}+H_{2}+J\mathbf{S}_{1}\cdot\mathbf{S}_{2}\label{dimer}%
\end{equation}
where $S_{1}=S_{2}=9/2.$ For each dimer there are $(2S_{1}+1)(2S_{2}+1)=100$
energy eigenstates. Like in a SMM Mn$_{4}$, each state can be labelled
approximately by two predominant quantum numbers $\left\vert m_{1}%
,m_{2}\right\rangle $ for two SMMs with $m_{1,2}=-9/2,-7/2,\cdots,9/2.$
Without the coupling $J$ the states $\left\vert m_{1},m_{2}\right\rangle $ and
$\left\vert m_{2},m_{1}\right\rangle $ are degenerated. As the two SMMs can be
regarded as truly identical particles there exists the permutation symmetry
between particle 1 and 2 and the eigenstates may possess parity symmetry. Thus
the eigenstates for the system are denoted by $\left\vert m_{1},m_{2}%
\right\rangle _{+}$ for even parity and $\left\vert m_{1},m_{2}\right\rangle
_{-}$ for odd parity. The antiferromagnetic coupling $J$ may remove the
degeneracy of these two states, but the parity in the states remain unchanged.
Even when we take into account the the coupling J and the transverse terms the
states become the linear combination of all possible states, for simplicity,
we still use the two dominant quantum numbers to represent the states. All the
energy eigenvalues by neglecting the local stray fields are plotted in Fig. 1.
The average over the local stray field does not move the position of energy
level crossing.

Before explaining the experimental observation from the dimer we first
consider the effect caused by the coupling $J$ between the two particles.
Assume the tunneling between the states $\left\vert m_{1}\right\rangle $ and
$\left\vert m_{1^{\prime}}\right\rangle $ under a sweeping field
$\mathbf{B(}=-c_{b}t)$ and the tunneling splitting between the two states is
$\Delta$, the pair of the splitting energy eigenvalues near the resonant point
can be written as
\begin{equation}
\varepsilon_{\pm}=\frac{1}{2}\left[  (m_{1}+m_{1^{\prime}})c_{b}t\pm
\sqrt{\left[  (m_{1}-m_{1^{\prime}})c_{b}t\right]  ^{2}+\Delta^{2}}\right]
\end{equation}
with$\ c_{b}=g\mu_{B}\mu_{0}\hbar dB/dt,$ and the two states are given by%
\begin{equation}
\phi_{\pm}^{1}(t)=\left(  \pm c_{\pm}\left\vert m_{1}\right\rangle +c_{\mp
}\left\vert m_{1^{\prime}}\right\rangle \right)  /\sqrt{2}%
\end{equation}
with $c_{\pm}=\sqrt{1\pm(m_{1}-m_{1^{\prime}})c_{b}t/\sqrt{\left[
(m_{1}-m_{1^{\prime}})c_{b}t\right]  ^{2}+\Delta^{2}}}.$ Before the resonant
tunneling, the initial state is at $\left\vert m_{1}\right\rangle ,$ i.e., at
$t=-\infty,$ $\phi_{+}(t)\rightarrow$ $\left\vert m_{1^{\prime}}\right\rangle
$ and $\phi_{-}(t)\rightarrow$ $\left\vert m_{1}\right\rangle ;$ after the
tunneling, at $t=+\infty,\phi_{+}(t)\rightarrow$ $\left\vert m_{1}%
\right\rangle $ and $\phi_{-}(t)\rightarrow-\left\vert m_{1^{\prime}%
}\right\rangle .$ When two identical particles are put together there are four
possible states: $\left\vert +,+\right\rangle _{+}=\phi_{+}^{1}\otimes\phi
_{+}^{2}$ with the energy $2\varepsilon_{+}$ $,$ $\left\vert +,-\right\rangle
_{\pm}$ $=\left(  \phi_{+}^{1}\otimes\phi_{-}^{2}\pm\phi_{-}^{1}\otimes
\phi_{+}^{2}\right)  /\sqrt{2}$ with the energy $\varepsilon_{+}%
+\varepsilon_{-},$ and $\left\vert -,-\right\rangle _{+}=\phi_{-}^{1}%
\otimes\phi_{-}^{2}$ with the energy $2\varepsilon_{-}.$ We denote the even
and odd parity of the states by the subscripts $\pm$, respectively. The
energies vary with time $t$, and are plotted in Fig. 2, for an illustration,
by choosing $m_{1}=m_{2}=-9/2$ and $m_{1^{\prime}}=m_{2^{\prime}}=7/2$ for the
model in Eq.(\ref{dimer}). The tunneling splitting between the two states
$\left\vert +,+\right\rangle _{+}$ and $\left\vert -,-\right\rangle _{+}$ is
$2\Delta$, the double of a single particle as expected. To see the coupling
effect of two particles, we plot the energy eigenvalues for several different
couplings in Fig. 2. It is obtained by the exact diagonalization of
$100\times100$ matrix for the Hamiltonian. The two states $\left\vert
+,-\right\rangle _{\pm}$ are degenerated for $J=0.0$. A very little coupling
$J$ may remove the degeneracy of the two states. The state $\left\vert
+,-\right\rangle _{-}$ has odd parity and does not take part in the tunneling
process as other three states have even parity. It is shown that the coupling
$J$ leads to two consequences: (1) The tunneling splitting from $\left\vert
-,-\right\rangle _{+}$ and $\left\vert +,+\right\rangle _{+}$ decreases very
quickly with increasing $J$, and almost closes for $J>0.3\times10^{-5}$K$.$ In
the dimer of [Mn$_{4}$]$_{2}$ the coupling $J\approx0.1$K and tends to
suppress the tunneling at this point completely. (2) The tunneling splitting
from $\left\vert -,-\right\rangle _{+}$ to $\left\vert +,+\right\rangle _{+}$
occurs at two separated points via a intermediate state $\left\vert
+,-\right\rangle _{+}$. The coupling $J$ provides an inner bias field to expel
the two resonant points away from the original ones of $\left\vert
+,+\right\rangle _{+}$ and $\left\vert +,-\right\rangle _{+}.$ This triangle
process reflects the structure of the tunneling of the two identical
particles. In the language of $m_{1}$ and $m_{2}$, the process from
$\left\vert m_{1},m_{1}\right\rangle _{+}$ to $\left\vert m_{1^{\prime}%
},m_{1^{\prime}}\right\rangle _{+}$ is described as follows: \textsl{the first
resonant tunneling occurs from }$\left\vert m_{1},m_{1}\right\rangle _{+}%
$\textsl{ to }$\left\vert m_{1},m_{1^{\prime}}\right\rangle _{+}$\textsl{, and
the second\ one follows from }$\left\vert m_{1},m_{1^{\prime}}\right\rangle
_{+}$\textsl{ to }$\left\vert m_{1^{\prime}},m_{1^{\prime}}\right\rangle
_{+}.$\textsl{ The explicit tunneling from }$\left\vert m_{1},m_{1}%
\right\rangle _{+}$\textsl{ to }$\left\vert m_{1^{\prime}},m_{1^{\prime}%
}\right\rangle _{+}$\textsl{ is suppressed completely by the coupling }$J$.%
\begin{figure}
[ptb]
\begin{center}
\includegraphics[
height=1.8792in,
width=2.6429in
]%
{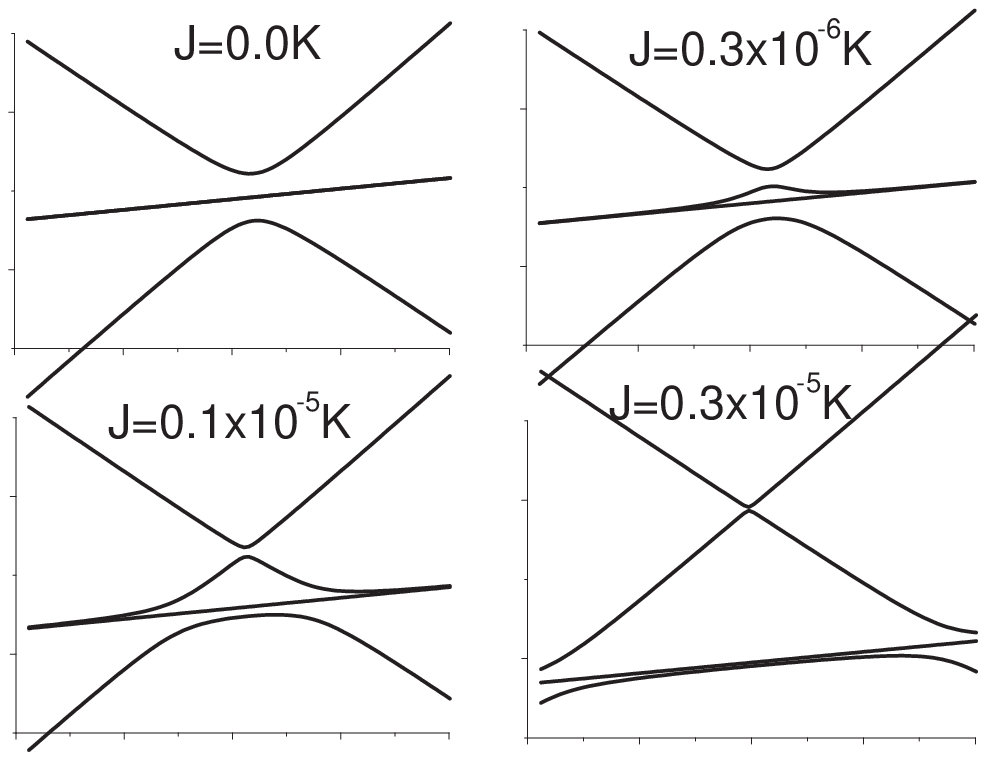}%
\caption{Energy levels of the states $\left\vert -9/2,-9/2\right\rangle _{+}$,
$\left\vert 7/2,7/2\right\rangle _{+}$ and $\left\vert -9/2,7/2\right\rangle
_{\pm}$ versus sweeping field at different coupling $J$ of two coupled
identical particles as an example to illustrate the tunnenling process from
$\left\vert m_{1},m_{1}\right\rangle _{+}$ to $\left\vert m_{1^{\prime}%
},m_{1^{\prime}}\right\rangle _{+}$ via $\left\vert m_{1},m_{1^{\prime}%
}\right\rangle _{+}.$ (x-axis: Energy/10$^{-4}$K, y-axis: Magnetic
Field/5$\times10^{-6}$T)}%
\end{center}
\end{figure}

Now we are ready to analyze the quantum tunneling in the dimer. Typical
hysteresis loops in magnetization versus sweeping external field applied along
the easy axis are observed. These loops display step-like features separated
by plateaus. The step heights become temperature-independent below $0.3$K, but
depend on the sweeping rate of magnetic field $c=dB/dt$. Derivatives of the
loops at different sweeping rate reflect that quantum tunneling occurs at
several points, but is absent at zero field. At high field the initial state
is $\left\vert -9/2,-9/2\right\rangle _{+}$, which has even parity. Due to the
permutation symmetry of identical particles all the tunneling to the states
$\left\vert m_{1},m_{2}\right\rangle _{-}$ with odd parity in this system are
prohibited. The tunneling process in the dimer can be understood essentially
by two triangle processes as shown in Fig. 1. Starting from the initial state
$\left\vert -9/2,-9/2\right\rangle _{+}$, the first level crossing happens at
magnetic field $b_{1}=-0.33$K, which is from $\left\vert
-9/2,-9/2\right\rangle _{+}$ to $\left\vert -9/2,9/2\right\rangle _{+}$ at
point 1, and the dual resonant point  is from $\left\vert
-9/2,9/2\right\rangle _{+}$ to $\left\vert 9/2,9/2\right\rangle _{+}$ at point
3 $b_{3}=+0.33$K in the first triangle process. The energy of the intermediate
state $\left\vert -9/2,9/2\right\rangle _{+}$ is independent of the external
field. The resonant field for the points 1 and 3 are $b_{1,3}=\pm9J/2g\mu
_{B}\mu_{0}$ from the model (Eq.(3))$.$ Thus J is calculated to be $0.1$K as
Wernsdorfer et al. did. The finite coupling does not lead to a tunneling
splitting at this point, which can be proved explicitly: for an integer n we
have
\begin{equation}
_{+}\left\langle -9/2,-9/2\right\vert H^{n}\left\vert -9/2,9/2\right\rangle
_{+}=0
\end{equation}
when the stray field is absent. The tunneling splitting at points 1 and 3 are
caused by the local stray field. These two points were consistent with
Ref.\cite{Wernsdorfer02}. The second process is from $\left\vert
-9/2,-9/2\right\rangle _{+}$ to $\left\vert 7/2,7/2\right\rangle _{+}$ via an
intermediate state $\left\vert -9/2,7/2\right\rangle _{+}.$ The two energy
level crossings are from $\left\vert -9/2,-9/2\right\rangle _{+}$ to
$\left\vert -9/2,7/2\right\rangle _{+}$ at point 2, and from $\left\vert
-9/2,7/2\right\rangle _{+}$ to $\left\vert 7/2,7/2\right\rangle _{+}$ at point
4. We take the parameters for D and E for a SMM Mn$_{4}$, and find that the
calculated resonant fields are $b_{2}=0.233$T and $b_{4}=0.861$T$,$ which is
very closed to the experimental data $0.87$T.\cite{Note1} As the points 2 and
3 are very close such that the resonant peaks are smeared to a broaden one. In
Wernsdorfer et al's paper they neglect the transition from $\left\vert
-9/2,7/2\right\rangle _{+}$ to $\left\vert 7/2,7/2\right\rangle _{+}$ and
thought that the third peak in the Fig. 4 of Ref.\cite{Wernsdorfer02} are
caused by those from $\left\vert -9/2,-9/2\right\rangle _{+}$ to $\left\vert
-9/2,5/2\right\rangle _{+}$ and from $\left\vert -9/2,9/2\right\rangle _{+}$
to $\left\vert 7/2,9/2\right\rangle _{+}$ (i.e. the point 5 in Fig. 1)$.$ $D$
was calculated to be $0.72$T by neglecting the transverse component $E$. After
a detailed analysis we found that the transverse component E remove the
degeneracy of $\left\vert -9/2,9/2\right\rangle _{+}$ and $\left\vert
-9/2,9/2\right\rangle _{-}.$ Even if we take $D=0.72$T we find that
$b_{2}=0.20$T, $b_{4}=0.833$T and $b_{5}=0.9138$T which is larger than
$0.87$T$.$ If we fix the point 5 at $0.87$T, $D$ is calculated to be $0.664$T
smaller than $0.72$T. The tunneling from $\left\vert -9/2,-9/2\right\rangle
_{+}$ to $\left\vert -9/2,5/2\right\rangle _{+}$ belongs to another the
triangle process and the splitting which is caused by the stray field is much
smaller than those at points 2 and 4. On the other hand we anticipate that the
weak coupling between two SMMs does not affect the intrinsic properties of a
SSM in the dimer too much. Our calculation shows that it is reasonable to take
the parameters $D$ and $E$ from the SMM Mn$_{4}$ for the model of the dimer of
the two SMMs with a weak coupling. It is worth pointing out that the coupling
J can also drive the tunneling splitting between some states such as
$\left\vert -9/2,+7/2\right\rangle _{+}$ and $\left\vert 9/2,7/2\right\rangle
_{+}$. However these tunnelings do not contribute significantly to what
observed in Ref.\cite{Wernsdorfer02} We do not discuss them here.

After determining the positions of the resonant points and model parameters we
are in a position to calculate the tunneling splitting, which determines the
transition rate in the Landau-Zener model. The exact diagonalization method is
applied to calculate the energy eigenvalues at different external field. The
sampling average is taken for the local stray field. For each sampling we
calculate the energy levels as in Fig. 1 and find the energy splitting
$\Delta_{n}$ at each resonant point. More than 1000 sampling are taken to
calculate the averaging tunneling splitting $\sqrt{\left\langle \Delta_{n}%
^{2}\right\rangle }$ for each distribution width $\sigma.$ The calculated
tunneling splittings are listed in Table I.

Table I: The calculated tunneling splitting $\sqrt{\left\langle \Delta
^{2}\right\rangle }$ in unit 10$^{-5}$K at different resonant points with
different distribution width $\sigma$ of the stray field $\mathbf{h}$ by using
the exact diagonalization method. ( $D=0.762$K, $E=0.0317$K.)%

\begin{tabular}
[c]{|l|l|l|l|l|l|}\hline
$\sigma/$T & 1 & 2 & 3 & 4 & 5\\\hline
0.000 &
$<$%
10$^{-7}$ & 2.21907 &
$<$%
10$^{-7}$ & 2.81552 & 1.52671\\\hline
0.010 & 0.01960 & 2.19227 & 0.01960 & 2.83557 & 1.59807\\\hline
0.020 & 0.03207 & 2.19487 & 0.03207 & 2.84268 & 1.59723\\\hline
0.035 & 0.04687 & 2.20155 & 0.04687 & 2.87249 & 1.61907\\\hline
0.050 & 0.06264 & 2.78338 & 0.06264 & 2.98306 & 1.68816\\\hline
\end{tabular}%
\begin{figure}
[ptb]
\begin{center}
\includegraphics[
height=1.9934in,
width=2.5858in
]%
{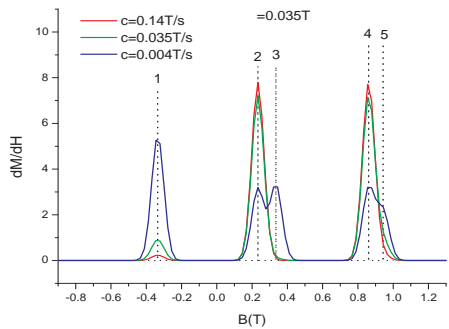}%
\caption{Calculated derivative of the hysterisis loop at different sweeping
fields based on the modified Landau-Zener model.}%
\end{center}
\end{figure}

The derivatives of the hysteresis loops at different sweeping rates in Fig. 4
of Ref.\cite{Wernsdorfer02} indicate that the peak heights in the derivatives
depend on the sweeping rate. The height of the first peak decreases with the
increasing rate while oppositely the second peak increases. This phenomenon
can be understood qualitatively in the modified Landau-Zener model. In
principle the time-evolution of the spin system can be reached by solving the
100($=(2S+1)(2S+1)$) coupled time-dependent Schr\"{o}dinger equation. As the
tunneling splitting is very small, the coupled equations near the two resonant
states can be reduced to an effective two-level system with the Hamiltonian%
\[
H_{eff}=\left(
\begin{array}
[c]{cc}%
(m_{1^{\prime}}+m_{2^{\prime}})(c_{b}t+h_{z}) & \Delta(\mathbf{h})/2\\
\Delta(\mathbf{h})/2 & (m_{1}+m_{2})(c_{b}t+h_{z})
\end{array}
\right)  .
\]
The tunneling splitting $\Delta(\mathbf{h})$ between two states $\left\vert
m_{1},m_{2}\right\rangle _{+}$ and $\left\vert m_{1^{\prime}},m_{2^{\prime}%
}\right\rangle _{+}$ can be obtained by diagonalizing the Hamiltonian with a
specific field $\mathbf{h}$. The state evolves with time,%
\[
\Phi_{eff}(t)=\exp\left[  -\frac{i}{\hbar}\int_{-\infty}^{t}H_{eff}%
(t)dt\right]  \Phi_{eff}(t=-\infty),
\]
and the magnetization varying with time is given \ by $M(t)=\left\langle
\Phi_{eff}\right\vert \mathbf{S}_{1}^{z}+\mathbf{S}_{2}^{z}\left\vert
\Phi_{eff}\right\rangle .$ The average over the stray field $\mathbf{h}$ is
taken for $\left\langle dM(t)/dt\right\rangle =\int d\mathbf{h}P(\mathbf{h}%
)dM(t)/dt.$ Physically, with the local stray field, the Landau-Zener
transition formula is given by%
\[
P_{LZ}=1-\left\langle \exp[-\pi\Delta_{mm^{\prime}}^{2}/\nu_{mm^{\prime}%
}]\right\rangle \approx\pi\left\langle \Delta_{mm^{\prime}}^{2}\right\rangle
/\nu_{mm^{\prime}},
\]
where $\nu_{mm^{\prime}}=2g\mu_{B}\hbar\left\vert \sum_{i=1,2}(m_{i^{\prime}%
}-m_{i})\right\vert dB/dt.$ The rate is proportional to the reverse of the
sweeping rate $c=dB/dt,$ approximately. The larger the sweeping rate is, the
less particles tunnel to the new state. The step height is related to the
transition rate by $\Delta M=P_{LZ}\sum_{i}(m_{i}-m_{i}^{\prime})$. The
presence of the local stray field will smear the \textquotedblleft
jump\textquotedblright\ of the magnetization around the resonant
point.\cite{Chen02} At a field $b$ around the resonant point, the variation of
the magnetization due to quantum tunneling can be approximately given by
$M(b)\simeq\Delta M\int_{-\infty}^{b}d\mathbf{h}_{z}P(\mathbf{h}_{z}),$which
leads to the derivative of the hysteresis loop around the resonant point,
$dM/db\simeq\Delta MP(b).$ The calculated results are plotted in Fig. 3.
Comparing with Fig. 4 in Ref.\cite{Wernsdorfer02}, we find that the numerical
results based a Landau-Zener model are consistent with the experimental
observation, and essentially reflect the sweeping rate effect on the peak
height of derivatives of hysteresis loops.

In conclusion, we study the coupling effect of two truly identical particles,
and analyze the quantum tunneling of magnetization in the supramolecular dimer
of two Mn$_{4}$s. The exchange coupling between two SMMs provides a biased
field to expel the tunneling\ to two new resonant points via an intermediate
state, and direct tunneling is prohibited. Based on the analysis we deduce the
model parameters from the experimental data, and find out that the coupling
does not change the model parameters for a SMM too much. Finally we point out
that the sweeping rate effect in the derivatives of hysteresis loops can be
explained quantitatively in the modified Landau-Zener model.

This work was supported by the Research Grant Committee of Hong Kong and a
CRCG grant of the University of Hong Kong.


\begin{thebibliography}{99}
\bibitem {high}As there exists higher order unknown transverse terms in the
model Hamiltonian, the tunnel splitting calculated in this paper is comparable
with the experimental data qualitatively.
\end{thebibliography}
\end{document}